\documentclass{mem}
\usepackage{natbib}
\usepackage{txfonts}
\usepackage{balance}
\usepackage{graphicx}
\usepackage[a4paper]{hyperref}
\idline{75}{282}

\newcommand{\rmkms}{\rm km s^{-1}}

\newcommand{\rmergs}{\rm erg s^{-1}}

\begin{document}

\title{
Mass of black holes:
}

\subtitle{The State of the Art}

\author{ B. \,Czerny\inst{1} 
\and M. Niko\l ajuk\inst{2} }

\offprints{B. Czerny}

\institute{
Copernicus Astronomical Center  --
Bartycka 18,
00-716 Warsaw, Poland \\
\email{bcz@camk.edu.pl}
\and
University of Bialystok, Faculty of Physics --
Lipowa 41, 
15-424 Bia\l ystok, Poland \\ 
\email{mrk@alpha.uwb.edu.pl}
}

\authorrunning{Czerny \& Niko\l ajuk}

\titlerunning{Mass of black holes}

\abstract{ 
In this small review we present the actual state the knowledge about
weighting black holes. Black holes can be found in stellar binary
systems in our Galaxy and in other nearby galaxies, in globular
clusters, which we can see in our and nearby galaxies, and in centres
of all well-developed galaxies. Range of values of their masses is
wide and cover about ten orders of magnitude (not taking into account
the hypothetic primordial black holes). Establishing the presence of
black holes, and in particular the measurement of their mass is one on
the key issues for many branches of astronomy, from stellar evolution
to cosmology.
              
\keywords{Stars: fundamental parameters --
galaxies: fundamental parameters -- galaxies: nuclei --
X-rays: binaries -- X-ray: galaxies}

}
\maketitle

\section{Introduction}

Astrophysical black holes (BH) are customarily divided into three classes: 
stellar mass black holes, with masses of a few up to twenty-thirty solar 
masses, intermediate mass black holes (IMBH) with masses of a few hundreds 
to a few thousands of solar masses, and massive (or super-massive) black 
holes residing in galactic dynamical centres, with masses from $10^5$ 
solar masses up.

Determination of values of black hole masses is important for the
three main reasons: (i) to distinguish small mass black holes from
neutron stars (ii) to analyse the luminosity states of accreting black
holes through the luminosity to the Eddington luminosity ratio (ii) to
study the growth of central black holes in galaxies. All three aspects
are important, the last one is a key issue in cosmology, in studies of
the galaxy evolution and AGN feedback.

The methods used for mass determination can be broadly divided into
the following three classes:
\begin{itemize}
\item {dynamical methods}
\item {spectra fitting methods}
\item {scaling methods}
\end{itemize}
although such a division is not necessarily unique. For example, the
reverberation method is rather an example of a dynamical method while
the Broad Line Region (BLR) radius vs. luminosity relation combined with a
line width fitting is somewhere in between the direct dynamical method
and a simple scaling law. In this review we will include it as a
scaling method.

There are several recent reviews covering the topic of black hole mass
determination
\citep[e.g.][]{Casares07,Ziolko08,Wandel08,VestergaardRev2009}. Here
we will not go into the details of the various methods and its caveats
but instead we will provide the collection of laws with short
description of applicability. The possibilities are broad so black
hole mass estimate is possible {\it nearly} for every object.
Different methods use observational data from various energy bands
from radio through NIR and optical to X-ray domain. Therefore, the
black hole mass measurement is one of the nice examples of the
multifrequency approach.

\section{Black hole mass determination methods}
\subsection{Dynamical methods}
\label{subsect:DynaMeth}

Dynamical methods are mostly based on the Keplerian motion of a {\it test
particle} around the black hole. The basis idea is simple: if for a
test particle we can measure the velocity $v$ at the circular orbit
and the radius $R$ of the orbit we obtain the value of the black hole
mass:
\begin{equation}
M = {v^2 R \over G}.
\end{equation}
The application is complicated by several factors. First, we have to
find a suitable {\it test particle}. Next, there is a problem of the
inclination of the orbit. Even for a circular orbit the measurement of
one of the velocity components does not allow for a unique
determination of the orbital speed.

We have a choice of the following {\it test particles} and measurements:
\begin{itemize}
\item {Companion star (in binary systems) -  mass function} 
\item {Companion black hole (e.g. OJ 287) - orbital period} 
\item {Nearby star (e.g. S02 star in Sgr A*) - apparent motion, radial
velocity}
\item {Nearby stars - stellar rotational velocity, stellar velocity 
dispersion and/or stellar luminosity profile}
\item {Broad Line Region clouds - line width and BLR radius from 
reverberation} 
\item {H2O masers - velocity and  radius from mapping} 
\end{itemize}

Of those methods the last one - water maser method - is the most
accurate since we observe water maser emission in an AGN only when
observing almost in the orbital plane of the emission.
Additionally, since the emission of several blobs located at various
distances from the black hole is measured we have a test whether the
motion is indeed Keplerian. The emission comes from the geometrically
thin disk, from distances of order of less than a fraction of a parsec
from the central black hole. Such measurement was done for NGC 4258
\citep{Miyoshi95,Herrn05}, and the most recent measurements
give the black hole mass value of $(3.82 \pm 0.01) \times 10^7
M_{\odot}$ water maser \citep[see][and the references
therein]{Siopis09}. Similar measurements were done for a few other
sources, but the water maser emission is rare (detected in about 10
percent of the nearby Seyfert 1.8 - Seyfert 2 galaxies,
\citealt{Kondratko06}), as it requires special inclination and
relatively low absorption as for a Seyfert 2 galaxy (highly inclined
AGN are extreme type 2 AGN), the data quality is usually not good
enough to allow for mass measurement (emission is unresolved), and in
some cases like in NGC 1068 the implied motion is clearly
non-Keplerian so the mass measurement cannot be performed reliably.
The method does not apply to Seyfert 1 galaxies or lower mass black
holes, and the majority of Seyfert 2 galaxies also do not show maser
action so the method has very limited applicability.

In case of stellar mass black holes, since they are generally in binary
systems we have a secondary playing the role of the {\it test
particle}. The companion star, however, is itself massive so the
measurement of the companion semi-amplitude of the radial
velocities, $K_{\rm c}$, and the period, $P$, of the orbital
motion results only in determination of the mass function
\begin{equation}
f(M) = {(M \sin i_{\rm c})^3 \over (M + M_{\rm c})^2} = {K_{\rm c}^3
P \over 2 \pi {\rm G}},
\end{equation}  
which gives only the lower limit for the black hole mass $M$. The
further determination of the black hole mass requires both an estimate
of the inclination of the orbit, $i_{\rm c}$, as well as the estimate of the
mass of the companion star, $M_c$. Best constraints are achieved if
multiwavelength measurements are used (e.g. to confirm the binary
period) and evolutionary constraints applied. New determinations of the
black holes were recently performed for example by
\cite{Orosz2009lmcx1} (LMC X-1, $M = 10.91 \pm 1.41 M_{\odot}$).

Overall, black hole masses in binary systems of the Milky Way were
determined for 24 objects (confirmed BH). The accurate values (error
less then 10\%) were obtained for 7 objects
\citep{Casares07,Ziolko08}.

Similar analysis of the binary motion was also used in the case of two
massive black holes. In OJ 287 the observed periodic outbursts every
12 years are interpreted as a result of a smaller massive black hole
($\sim 1.3 \times 10^8 M_{\odot}$ crossing the accretion disk of a
more massive ($\sim 1.8 \times 10^{10} M_{\odot}$) black hole (see
\citealt{Valtonen09} and the references therein). However, this object
is exceptional, with the optical data covering more than 100
years. The search for other massive binary black hole continues but
the list of candidates is short (see e.g. \citealt{Komossa09}) and the
available data do not allow for black hole mass measurement at the
basis of binarity itself.

Single star as a {\it test particle} can also be used in the case of
the massive black hole mass measurement for Sgr A*. In this case the
star is indeed a test particle, but stellar orbits are elliptical and
inclined. The peculiar motion of these young bright stars can be
followed using the active optics in the IR band. One of the stars
(S02) already completed the whole orbit and the most recent
determination of the Sgr A* mass based on that result is 
$(4.1 \pm 0.6)\times 10^6 M_{\odot}$ \citep{Ghez08}. 
Similar, and even more accurate
value is obtained if all 28 monitored stars are used ($(4.31 \pm 0.06)
\times 10^6 M_{\odot}$, \citealt{Gillessen09}).  The provided
uncertainty of the measurement is additionally enhanced to $0.36 \times 10^6
M_{\odot}$ due to systematic error in the estimate of the distance
to the Galactic centre.

The method based on a single star cannot be used now for other, even
nearby galaxies or for centres of the globular clusters (due to the
crowded fields).

Thus the next approach is to use the measurement of the rotational
velocity or/and velocity dispersion of the stellar distribution to
infer the presence of the central black hole. The method is not a
simple one since it requires modelling of the stellar motion on the
top of the measurement of the stellar velocity from the width of
absorption lines. If the measured stars are not close enough so that
the gravitational potential contains the contribution from the stars
themselves the problem is particularly difficult.

The black hole mass measurement for nearby non-active black holes
based on stellar dispersion was performed for several objects
observed by HST or ground-based telescopes, preferably with active optics
\citep{Magorrian98,Gebha2000,FerraMerritt2000,Krajno2009}.
As was shown for example by 
\citet{Gultekin09a} for NGC 3607, the fit with a central black hole of 
mass of $(1.2 \pm 0.4) \times 10^8 M_{\odot}$ is better by $\chi^2 =10.6$ than
the fit without a central black hole.
Application of the method to globular clusters in our Galaxy is
even more difficult since the size of the region dominated by the
central black hole has smaller angular size. Minor changes about the
assumption of the stellar dynamics change the result drastically. For
example, the fit to the globular cluster $\omega$ Centauri gives no black
hole for core assumption for the stellar distribution but for the cusp
profile assumption it gives the black hole mass of $(8.7 \pm 2.9)
\times 10^3 M_{\odot}$ \citep{vanderMarel09}.
Another example of the potential problems was shown by
\citet{Fregeau09}: the inclusion of natal kicks of white dwarfs in globular
cluster NGC 6397 removes the need for the IMBH at the cluster
centre. 

The determination of the black hole mass directly from the
stellar dispersion velocity profile should not be confused with a
global measurement of bulge velocity dispersion and black hole mass
determination based on the scaling between the two since the motion of
the stars in bulge is not determined dynamically by the black hole
mass; this method will be described in Sect.~\ref{subsec:scaling}.

Dynamical methods can be extended to distant objects for active
galactic nuclei since in this case the gas of the Broad Line Region
(BLR) can play the role of the {\it test particle}. The BLR is never
resolved but the emission line profile provides the information on the
velocity and the {\it orbital radius} can be measured from the time
delays between the variations of the continuum and the response of the
line, i.e. from reverberation (see e.g. \citealt{Peterson04} for the
general description).

The line profile is converted to the velocity either through
measurement of the full-width-at-half maximum (FWHM), or through
line dispersion (second moment of the line profile), $v_{\sigma}$. 
As argued by \cite{Denney2009}, the first method is better for
low signal to noise spectra while the second method is more accurate 
for higher signal to noise spectra since it is then less sensitive to
details of the subtraction of the narrow component of the line profile. 

The black hole mass is then determined from the relation:
\begin{equation}
M_{\rm BH} = f_1 {R v_{\rm FWHM}^2 \over \rm G}
\label{eq:MbhFWHM}
\end{equation}
or from the relation:
\begin{equation}
M_{\rm BH} = f_2 {R v_{\sigma}^2 \over \rm G}.
\label{eq:Mbhsigma}
\end{equation}

The problem of the method is the geometry of the BLR. If we assume the
spherical geometry, as introduced by \citet{Netzer90}, the coefficient
$f= {3 \over 4}$ in the formula above is determined
uniquely. This approach was used  by \citet{Wandel99, Kaspi2000}.

However, the BLR is likely to be flattened 
 \citep[see e.g.][]{McLure01} 
so the coefficient was later determined at the basis of the comparison
between two mass determination methods (reverberation and stellar
dispersion, \citealt{Onken2004}) and this coefficient was adopted in
Eqs.~(\ref{eq:MbhFWHM}) and (\ref{eq:Mbhsigma}). It is by a factor $\sim
1.8$ higher than in spherically symmetric case. 
In addition, \citet{Collin2006} reported that the value of the
coefficient $f$ is dependent on type of AGN (for example $f_1$ in
Eq.~(\ref{eq:MbhFWHM}) for Seyfert 1 galaxies with broad emission lines
is less than in case of Narrow Line Seyfert 1 (NLS1) galaxies).

Reverberation is time consuming but nevertheless it was performed for
over 40 objects \citep[see][and the references
therein]{PeterFera2004,Bentz2009a,Bentz2009c}. It includes several
Seyfert 1 galaxies, about 15 quasars and 7 NLS1 galaxies. 

The problem in this method is the potential dependence of the scaling
factor on the inclination angle. The issue was discussed in a number
of papers \citep[e.g.][]{Collin2006,niko2006}, and the systematic
error even by a factor of 3 can arise due to this effect in some
cases (see discussion by \citet{Krolik_blr}).

Additionally, \citet{Marconi2009} suggested that the effect of
radiation pressure should be included in determination of the black
hole mass based on dynamical methods (i.e. virial theorem). However,
\cite{onken_pressure} argued against the importance of this effect.

\subsection{Spectra fitting methods}
\label{sect:spectral_fitting}

The spectra fitting methods are based on the assumption that we have a
reliable model of the emission coming from the vicinity of a black
hole. Such a model usually depends on the black hole mass, but also on
some other parameters, mostly accretion rate, but also the spin of the
black hole and the viewing angle of the nucleus. Therefore, using
those methods is rather difficult and the results should be taken with
much care.

In high accretion rate sources, like Galactic black holes in their
soft states or Narrow Line Seyfert 1 galaxies and quasars, the
emission is mostly dominated by the emission of an accretion disk seen
in X-rays (GBH) or optical/UV (AGN). In exceptional AGN the peak
emission is also seen in X-rays (RE J1034+396, \citealt{SoriaPuch02}).

If we have a single measurement of the monochromatic flux at
$\nu_0$ (in the power-law region of the big blue bump)  and the
position of the maximum of the disk component spectrum on $\nu
F_{\nu}$ diagram, $\nu_{\rm max}$, then assuming a non-rotating black
hole, disk inclined at 60 deg and extending to the innermost stable
circular orbit (ISCO) and a local black body without any correction
connected with the colour temperature we can have a black hole mass
estimate:
\begin{eqnarray}
\log M_{\rm BH} = {1 \over 2} \log L_{\nu_0} - {1 \over 6} \log \nu_0 +
\nonumber \\
- {4 \over 3} \log \nu_{\rm max} + 16.515
\end{eqnarray}
\citep{Tripp94}. This can be applied to bright quasars if only
broad band low quality spectrum is available. In Seyfert galaxies the
spectral slopes are much softer than expected from the Shakura-Sunyaev
disk emission (i.e. the effects of X-ray irradiation, extinction and
outflow are important) so the formula is inadequate. For higher
quality broad band spectra more advanced models can be fitted (see
e.g. \citealt{Davis07} based on \citealt{Hubeny2000} models;
irradiation effect in \citealt{Loska04} and
\citealt{Cackett07,bczjan07}).

For galactic sources, more advanced disk spectra model are used
\citep[e.g.][]{Sadek09} and with the knowledge of the black hole
mass from dynamical method (binary motion) they are used instead to
infer the black hole spin \citep[e.g.][]{McClintock06}.

However, in order to rely on such mass determination we must be sure
that the measured emission does come from a fairly standard accretion
disk. In some sources, like ULX, this is highly uncertain. For
example, thermal disk + power law fits to X-ray spectra imply IMBH
\citep[e.g.][]{Miller04} while other decomposition of the
spectra implies the ultraluminous high states of 100 $M_{\odot}$ BH 
\citep[e.g. ][]{DoneKubota06,Roberts07}. Recent study of the sample 
of 94 Chandra sources \citep{Berghea2008} shows that the cool disk
spectral component is unrelated to luminosity which implies that
indeed we do not understand the spectra! Arguments against low masses
in ULX sources (i.e. against strong beaming) are rather indirect
(see the discussion of the sources ULX-1 ESO 243-49 by \citep{Farell2009}
 for a recent example of such analysis). Therefore, conclusions on the
presence or absence of IMBH based on spectral analysis cannot be done.

Accreting black holes also show the presence of the X-ray emission and
the reflection component coming from X-ray reprocessing by an
accretion disk. The iron line formed in this process can be also used
for measurement of the black hole parameters. The line profile is mostly
sensitive to the black hole spin but if X-ray time variability is
included then iron line reverberation also should allow (in principle,
when next generation large area X-ray telescope will operate) for
black hole mass measurement \citep{Dovciak04}.

\subsection{Scaling methods}
\label{subsec:scaling}

Scaling methods are invaluable for black hole mass determinations in
large AGN surveys as well as when a simple black hole mass estimate is
needed for a specific object. There are numerous methods based on
various observables. The accuracy of the methods is difficult to
assess for a single object so two or more methods should be used
whenever possible. These methods are based on scaling of measured
properties with the mass found earlier by independent mass and the
discuss quantity measurement.

Most broadly used nowadays is the method of the black hole mass
measurement from a single spectrum in IR, optical or UV band. The
method originally comes from reverberation approach but is based on
empirical connection established between the BLR radius and the source
monochromatic luminosity \citep[see e.g.][]{Wandel99,Kaspi2000}.
If the starlight contamination is likely to be unimportant,
or it is accounted for, the best relation to use is
\begin{eqnarray}
\log \frac{R_{\rm BLR}}{1 \rm \ lt\ day} = -21.3 + 
\nonumber \\
+ (0.519^{+0.063}_{-0.066}) \log(\lambda L_{\lambda} (5100\AA))
\end{eqnarray}
(based on HST spectra, \citealt{Bentz2009a}) and consequently the
formula for the black hole mass determination based on
$\ion{H}{\beta}$ line reads:
\begin{eqnarray}
\log M_{\rm BH} = \log \Bigg[ \Big( 
\frac{\rm FWHM( \ion{H}{\beta}) }{1000 \, \rmkms} \Big)^2 \times
\nonumber \\ 
\times \Big( \frac{ \lambda L_{\lambda}(5100\AA)}{10^{44} \rmergs} 
\Big)^{0.519} \Bigg] + \log f_1 + 6.81 \ ,
\end{eqnarray}
{where $f_1$ is the factor from Eq.~(\ref{eq:MbhFWHM}) and $M_{\rm BH}$
is in units of $M_{\odot}$. }
If the starlight contamination is likely,
the older version of the phenomenological fit may be more accurate in
practice. 
\citet{Kaspi05} give the relation
\begin{eqnarray}
\frac{R_{\rm BLR}}{10 \ \rm lt\ days} = 2.39
\Bigg( \frac{\lambda L_{\lambda}(5100\AA)}{10^{44} \rmergs}
\Bigg)^{0.67 \pm 0.05}
\end{eqnarray}
and from this it follows that
\begin{eqnarray}
\log M_{\rm BH} = \log \Bigg[ \Big( 
\frac{\rm FWHM( \ion{H}{\beta}) }{1000 \, \rmkms} \Big)^2 \times
\nonumber \\ 
\times \Big( \frac{ \lambda L_{\lambda}(5100 \AA)}{10^{44} \rmergs} 
\Big)^{0.67} \Bigg] + \log f_1 + 6.66 \ .
\end{eqnarray}

In their AGN analysis \citet{VesterPeter06} use two formulae:
\begin{eqnarray}
\log M_{\rm BH} = \log \Bigg[ \Big( 
\frac{\rm FWHM( \ion{H}{\beta}) }{1000 \, \rmkms} \Big)^2 \times
\nonumber \\ 
\times \Big( \frac{ \lambda L_{\lambda}(5100 \AA)}{10^{44} \rmergs} 
\Big)^{0.5} \Bigg] + (6.91 \pm 0.02) \ ,
\end{eqnarray} 
or
\begin{eqnarray}
\log M_{\rm BH} = \log \Bigg[ \Big( 
\frac{\rm FWHM( \ion{H}{\beta}) }{1000 \, \rmkms} \Big)^2 \times
\nonumber \\ 
\times \Big( \frac{L(\ion{H}{\beta})}{10^{44} \rmergs} \Big)^{0.5} \Bigg] 
+ (6.67 \pm 0.03) \ .
\end{eqnarray} 

If $H\beta$ is not available, other lines can be used. For example,
the recent formula based on the $\ion{Mg}{ii}$ line reads:
\begin{eqnarray}
\log M_{\rm BH} =  \log \Bigg[ \Big( 
\frac{{\rm FWHM}( \ion{Mg}{ii}) }{1000 \, \rmkms} \Big)^2 \times
\nonumber \\ 
\times \Big( \frac{\lambda L_{\lambda}}{10^{44} \rmergs} \Big)^{0.5} \Bigg] 
+ zp(\lambda) \ ,
\end{eqnarray} 
where $zp(\lambda)$ = 6.72, 6.79, 6.86 and 6.96 for $\lambda 1350$\AA,
$\lambda 2100$\AA, $\lambda 3000$\AA, and $\lambda 5100$\AA, respectively
\citep{VesterOsmer09}. There is also a version of this formula based
on the line flux instead of the continuum monochromatic flux
 \citep[see e.g.][]{Kong06}.

If even $\ion{Mg}{ii}$ is outside the measured spectrum, there is 
a possibility to use $\ion{C}{iv}$ line \citep{VesterPeter06}:
\begin{eqnarray}
\log M_{\rm BH} = \log \Bigg[ \Big( 
\frac{\rm FWHM( \ion{C}{iv}) }{1000 \, \rmkms} \Big)^2 \times
\nonumber \\ 
\times \Big( \frac{ \lambda L_{\lambda}(1350 \AA)}{10^{44} \rmergs} 
\Big)^{0.53} \Bigg] + (6.66 \pm 0.01) \ ,
\end{eqnarray} 

This line is perhaps less reliable as belonging to High Ionisation
Lines (HIL) and being more likely influenced by the radiation pressure
that Low Ionisation Lines (LIL) (like $\ion{H}{\beta}$ and
$\ion{Mg}{ii}$).

Those methods work for type 1 AGN, i.e. Seyfert 1 galaxies and
quasars, in Seyfert 2 galaxies we do not see the BLR region. However,
in those sources we can use scaling based on the bulge properties
which generally seem to apply for all AGNs well as non-active
galaxies.

The best relation, with very low dispersion, was found between the
stellar dispersion in the bulge and the black hole mass 
\citep{Gebha2000,FerraMerritt2000}. \citet{Gultekin09b}
based on observations of over 49 observations of galaxies give
formula: 
\begin{eqnarray}
\log M_{\rm BH} = 8.12 + 
\log \Big( \frac{\sigma}{200 \, \rmkms} \Big)^{4.24 \pm 0.41} \ ,
\label{eq:MbhSigma}
\end{eqnarray} 
which implies the surprising relation between the black hole mass and
the bulge mass \citep{KormRich95,Magorrian98}. If the stellar
dispersion in the bulge is not measured we can use first the relation
between the bulge luminosity and the bulge mass 
\begin{eqnarray}
\log \Big(\frac{M_{\rm bulge}}{M_{\odot}} \Big) = -1.11 +
1.18 \log \Big( \frac{L_{\rm bulge}}{L_{\odot}} \Big)
\end{eqnarray} 
Then we can combine it with the black hole mass vs. bulge mass
relation in order to receive $M_{\rm BH}$ \citep{HaringRix2004}: 
\begin{eqnarray}
M_{\rm BH} = 1.58 \times 10^8  \, \Big( \frac{M_{\rm bulge}}{10^{11}
M_{\odot}} \Big)^{1.12 \pm 0.06} M_{\odot} \ .
\end{eqnarray} 
If the bulge luminosity is measured in B band then we have to
recalculate it to V band through empirical relation $B-V=0.8$
\citep[e.g.][]{BianZhao2003} or directly use the formula proposed by
\citet{KormGebh01}
\begin{equation}
M_{\rm BH} = 0.78 \times 10^8 \, \Big( \frac{L_{\rm B,bulge}}{10^{10}
L_{\rm B, \odot}} \Big)^{1.08} M_{\odot} \ .
\end{equation}

For AGN the most recent scaling relation found by \citet{Bentz2009b} is:
\begin{equation}
\log M_{\rm BH} = 7.98 + \log \Big( {L_{\rm V,bulge} \over 10^{10}
L_{\odot}} \Big)^{0.80 \pm 0.09}
\end{equation}

For highly inclined radio galaxies the following relation between host 
galaxy absolute
magnitude at R-band, $M_{\rm R}$ and $M_{\rm BH}$ can be applied
\begin{equation}
\log M_{\rm BH} = -0.5 M_{\rm R} - 2.74,
\end{equation}
\citep{mclure02} where $M_{\rm R}$ is the absolute R-Cousins bulge luminosity.

For BL Lac objects the similar relation, but with a different shift,  
was proposed by
\citet{Bettoni2003}:
\begin{equation}
\log M_{\rm BH} = -0.50 M_{\rm R} - 3.00 \ .
\end{equation}
More complex method of deriving the black hole mass from the
Two-Micron All-Sky Survey K-band bulge luminosity was presented by
\citet{Vasudevan}.

Finally, 278 radio-loud AGNs (including 146 BL Lac sources) were observed by 
\citet{ZhouCao2009}. A significant correlation was found between 
the Lorentz factor of the jet, $\gamma_{\rm min}$, and black hole mass
which opens a possibility to estimate the last quantity:
\begin{equation}
\log M_{\rm BH} = 3.26 \log \gamma_{\rm min} + 5.81 \ .
\end{equation}

In case of AGN we have also a possibility to use the profile of
optical [$\ion{O}{iii}$] line (from NLR) as a proxy for the stellar
dispersion velocity as advocated by several authors and use the
Eq.~(\ref{eq:MbhSigma}). The accuracy is improved
with the use of a relation
\begin{equation}
\sigma_* = \rm FWHM([\ion{O}{iii})/2.35
\end{equation}
\citep{Gaskell_oiii}. This possibility is particularly interesting for radio
galaxies where obscuration of the central region is considerable due
to the host galaxy when the source is seen in the host galaxy plane,
as it is in CSO sources. When mid-IR spectrum is available, other NLR
lines can be used, like [$\ion{Ne}{v}$], [$\ion{O}{iv}$] \citep{kalliopi2008}.

Finally, also X-ray data provide us with the black hole mass
measurement possibility. One of the advantages of those methods is the
expected independence on the inclination effects since the X-ray
emission is rather isotropic.

X-ray lightcurves of accreting black holes are dominated by aperiodic
red noise and power spectral densities (PSDs) are well represented
by several Lorentzians. Such analysis is mainly done using high quality 
X-ray data  for X-ray
Binaries in our Galaxy \citep[see e.g.][for Cyg X-1, GX
339-4]{Nowak2000,Pottschm2003}. In case of low quality X-ray data,
generally in AGNs and ULXs, PSD of those sources are represented by a
power law with one or two breaks \citep{Uttley2002,Heil2009}. However,
recently \citet{McHardy07} reported finding 2 Lorentzian components
in PSD of nearby NLS1 galaxy Ark 564.

The overall shape of the power spectrum depends on the luminosity
state (the effect well studied in galactic black holes) and shifts
linearly with the black hole mass.  High frequency break was suggested
to scale with mass in AGN \citep{Papadakis04} as
\begin{equation}
M_{\rm BH} = 10^7 {1.7 \times 10^{-6} \over f_{\rm hfb}} M_{\odot},
\end{equation}
where $f_{\rm hfb}$ is the high frequency break in
Hz. \citet{McHardy06} suggest a formula which includes also the
luminosity term:
\begin{eqnarray}
\log f_{\rm hfb} [{\rm day}] = -2.17 \log \Big(
\frac{M_{\rm BH}}{10^6 M_{\odot}} \Big) + 
\nonumber \\
+ 0.90 \log \Big( \frac{L_{\rm bol}}{10^{44} \rm erg s^{-1}} \Big) +
2.42 \ . 
\end{eqnarray}

However, the form and strength of the luminosity dependence is an open
issue and likely depends of the AGN type (Seyfert 1 vs. NLS1 vs. LLAGN). 
In galactic sources the low frequency part of the
X-ray power spectrum clearly shows a dependence on the hard or soft state of 
the source. 
No strong variability of the high frequency part of
the X-ray power spectrum is seen for galactic sources in their hard
states \citep{Gier08}. Similar disconnection with luminosity and
dependence only on $M_{\rm BH}$ is seen in Sgr A* variability observed in
Near IR \citep{Meyer09}: the break in the power spectrum of the NIR 
lightcurve fits expectations from AGN despite the huge difference in 
the bolometric luminosity between AGN and Sgr A*.

An interesting version of this scaling emerged from the approach of
\citet{Hayashida98} and uses only the high frequency part of the
power spectrum, or, in practice, the X-ray excess variance measured in
the timescales shorter than the timescale corresponding to $1/f_{\rm
hfb} $ in previous formulae. The relation
\begin{equation}
M_{\rm BH} = 1.92 {T - \delta t \over \sigma^2_{\rm rms}} M_{\odot}
\end{equation}
\citep{niko2006,guryn2009} is based on the scaling with Cyg X-1 
and uses 20 $M_{\odot}$ for Cyg X-1 mass; study for a number of
galactic sources gives somewhat lower coefficient of 1.24
\citep{Gier08}. Here $T$ is the duration of the lightsurve and
$\delta t$ is the time bin, both in seconds, and the X-ray excess
variance, $\sigma^2_{\rm rms}$ is dimensionless, normalized by the
average flux.

In addition to broad band power spectra, quasi-periodic oscillations
(QPO) are seen occasionally in X-ray Binaries and ULXs. Their use for
black hole mass measurement is complicated by the fact that there are
several types of QPO, and firm identification of the QPO type is
needed to infer the black hole mass. QPO specific example of the
application can be found in \citet{Shapka07} who derived the black
hole mass in Cyg X-1 of $8.7 \pm 0.8 M_{\odot}$ from the black hole
mass in GRO J1655-40 at the basis of their QPO scaling with mass and
X-ray spectral slope. There were several claims about the presence of
QPO also in AGN, and recently \citet{Gier08a} reported about discovery
of an unmistakable QPO in NLS1 RE J1034+396. \citet{MiddleDone09} have
estimated the mass of a black hole in this object ($2-3 \times 10^6
M_{\odot}$).

However, caution in the case of X-ray methods is also needed since
some weak inclination effect can be expected due to the general
relativity causing in general slight enhancement in variability for
highly inclined objects seen in theoretical studies
\citep[e.g.][]{AbramBao94,bcz2004}.

\section{Consistency checks}

Since all the methods may contain systematic errors the best test of the
accuracy of the black hole mass determinations is to use two or more 
independent methods. This is particularly important for all the scaling 
approaches. In the case of AGN, also black hole mass determinations done at 
various epochs are interesting since they rely on the spectra which vary in 
time. Here we list a number of such comparisons.

A test of the reliability of a single epoch mass measurement based on
the BLR radius-luminosity scaling (see Sect.~\ref{subsec:scaling}) was
performed using multi epoch measurements for two galaxies (NGC 5548
and PG 1229+204, \citealt{Denney2009}). The dispersion of the mass
measurement due to the source variability was less than 0.1 dex for
high signal to noise spectra (above 20).

The use of various optical/UV lines is nicely illustrated in
\citet{VestergaardRev2009} (see her Fig.~3). There are visible 
discontinuities in the mass distribution with redshift due to the
forced change from $\ion{H}{\beta}$ to $\ion{Mg}{ii}$ and finally
$\ion{C}{iv}$, but less than a factor 2. \citet{Risa09} confirm
previously suggestions \citep[see e.g.][]{Shen08} that $\ion{C}{iv}$
line is not a reliable indicator of BH masses. Much better is to use
$\ion{Mg}{ii}$. However, BH masses determined from this ion has
systematic error so the $\ion{Mg}{ii}$ mass can be corrected using the
relation $\log [M_{\rm BH}(\ion{H}{\beta})] = 1.8\times \log [M_{\rm
BH}(\ion{Mg}{ii})] - 6.8$.

The comparison of two methods was done for a water maser source NGC
4258, with the black hole mass of $(3.82 \pm 0.01) \times 10^7
M_{\odot}$ from water maser \citep{Herrn05} and $(3.3 \pm 0.2) \times
10^7 M_{\odot}$ from stellar dispersion \citep{Siopis09}. Stellar
dispersion against reverberation was also tested for NGC 4151, stellar
dispersion gives $(4 - 5) \times 10^7 M_{\odot}$ \citep{Onken2007} and
reverberation gives $4.57^{+0.57}_{-0.47} \times 10^7 M_{\odot}$
\citep{Bentz2006}.

Based on such tests, \citet{VestergaardRev2009} estimates that the
systematic error in secondary estimates (i.e. scaling laws) is likely
to be from 0.3 dex for stellar dispersion in the bulge to 0.7 dex when
using bulge luminosity or NLR line width.

Interesting tests of the effects of inclination in AGN were performed
by comparison of the reverberation method and the X-ray excess
variance \citep{niko2006}. The change of the inclination from
$\sim 15 $ deg to $\sim 60$ deg is likely connected with the change in
the factor $f$ in Eq.~\ref{eq:MbhFWHM} by a factor 4 although errors in this
analysis are large.

\section{Summary of the results}

\subsection{Galactic black holes}

The typical accuracy in the determination of the black hole masses is
about a factor 2 or better. The best known black hole mass
(microquasar GRO J1655-40, $M_{BH} = 6.3 \pm 0.5$,
\citealt{Greene2001}) was achieved due to the dynamical method
(accurate period determination, inclination from modelling of the
lightcurve due to the companion ellipticity, mass ratio also from
modelling the ellipticity). Overall, there are 
over 40 (24 confirmed BH)  
determinations of the black hole mass in
such systems in our Galaxy and a few in the binaries in nearby
galaxies like in LMC and M33 (e.g. eclipsing binary M33 X-7, with
black hole mass of $15.65
\pm 1.45 M_{\odot}$; \citealt{Orosz07}). 

\subsection{Intermediate mass black holes}

This is the hottest subject in the Art of black hole mass
measurement. First, the exact range is not specified, so it is not
clear whether relatively small but otherwise typical AGN (i.e. located
at the centres of their host galaxies) black holes from the range
$\sim 10^5 - 10^6 M_{\odot}$ (see sample of \citealt{Desroches09})
belong to this class. Their existence does not cause much doubt
(e.g. \citealt{HoRev}). The
on-going stellar dispersion measurements in hundreds of near-by galaxies
(see for example \citealt{Ho2009}) may bring more of them although measuring
black holes with masses below $10^6 M_{\odot}$ is very difficult.

More questionable is the existence of the black holes with the masses
of hundreds-thousands of the solar mass.
Determination of $M_{\rm BH}$ in several ULX sources, which are hosted
by galaxies taken from  Messier and NGC catalogues, and the discussion 
of the stellar mass versus IMBH interpretation
can be found in \citet{Zamp09} (see also references therein).

\subsection{Non-Active galaxies or Weakly Active galaxies}

The best example is of course the Milky Way galaxy, with the recent
mass determination (see Sect.~\ref{subsect:DynaMeth}). The mass in
other nearby galaxies is known less accurately (for example, triple
nucleus of Andromeda hosts a black hole mass in P3 of $(1.1 - 2.3) \times 10^8
M_{\odot}$, \citealt{bender05})
but mass estimate (usually through stellar dispersion measurement) is available for hundreds of objects \citep{HoRev}.

\subsection{Active Galactic Nuclei}

The number of known AGN 
is of order of $10^5$ (in SDSS DR7) 
and rising due
to massive surveys. The best results are obtained from water maser discussed
in Sect.~\ref{subsect:DynaMeth}. Generally, mass determination for 
tens/hundreds/thousands of objects were performed by several authors
\citep[e.g.][]{WooUrry02,VesterPeter06,Vestergaard08,Shen08,Fine08,kelly09,wu09}. 

Numerous large surveys supplemented with the black hole mass and the
Eddington ratio determinations \citep[e.g.][]{Hickox2009} the
available scaling laws from the list discussed in
Sect.~\ref{subsec:scaling}. The Eddington ratio is the frequently
determine using another scaling to derive the bolometric luminosity,
for example from the optical monochromatic flux. 
\citet{Richards06} (see their Fig.~12) analysed spectra energy
distributions of quasar type 1 and gave a formula:
\begin{equation}
L_{\rm bol} = {\rm BC}_{\lambda} \times \lambda L_{\lambda} \ ,
\end{equation}
where the bolometric corrections BC$_{\lambda}$ = 3.81, 5.15, 9.26 are
for measurements of luminosity, $\lambda L_{\lambda}$, made at
$\lambda$= 1350, 3000 and 5100 \AA, respectively.

It is also possible to use the near-IR monochromatic luminosity
\begin{equation}
\label{eq:bolo_IR}
L_{\rm bol} = 6.4 L(\rm NIR) 
\end{equation}
\citep{cao05}
or from the broad band X-ray luminosity \citep{Hopkins2007}
\begin{equation}
\label{eq:bolo_X}
L_{\rm bol} = (10-20) L_{\rm X}(0.5-8 {\rm keV}).
\end{equation}

More complex, luminosity-dependent corrections can be found in
\citep{Hopkins2007}. \citet{Hickox2009} used the coefficient in
Eq.~(\ref{eq:bolo_IR}) by a factor of two lower than derived by
\citep{Hopkins2007} in order to have consistent results for AGN with
both IR and X-ray measurements available.

In heavily obscured Seyfert 2 galaxies or radio galaxies the
bolometric luminosity can be estimated from the [$\ion{O}{iii}$] line
luminosity:
\begin{equation}
L_{\rm bol} = {\rm C} L([\ion{O}{iii}]),
\end{equation}
where ${\rm C} = 87, 142 $ and $454$ for logarithmic luminosity ranges
of [$\ion{O}{iii}$] line 38 - 40, 40 - 42 and 42 - 44, correspondingly
\citep{lamastra}.

\section{Discussion}

The large number of various methods allow to estimate the black hole
mass in most astronomical objects quite reliably, i.e. within a factor
of a few. Accurate determination of the error of a particular
measurement is difficult since the errors are mostly systematic.

However, there are still classes of objects in high need for better
mass determination. Those are first of all black holes in ULX and in globular
clusters. The existence of IMBH is an interesting question both from the point 
of view of the accretion pattern (stellar mass black hole hypothesis for ULX 
sources requires very high Eddington ratios and strong beaming). Also some 
other types of active nuclei like weak line quasars (WLQ) and radio-loud 
objects like BL Lacs are still in a need for mass determinations although
some estimates are possible even in those difficult cases 
\citep[see e.g.][]{Hry2009}.

The mass determination together with the estimate of the bolometric
luminosity opens a possibility of statistical studies in case of
massive black holes since thousands of objects are available. Such a
black hole spends most of the time in a relatively quiet
state. Eddington ratios cover the range from $10^{-9}$ to $\sim 1$,
with LINERS (and Seyfert 2 galaxies) grouped at $\sim 10^{-4}$
\citep{HoRev}, Seyfert 1 galaxies at $\sim 10^{-2}$, and quasars as well as 
Narrow Line Seyfert 1 galaxies close to 1. Future evolutionary studies should 
explain the observed ratios of these types of galaxies. 


\begin{acknowledgements}
We are grateful to Joasia Kuraszkiewicz and Amri Wandel for helpful
discussions. Part of this work was supported by the grant NN 203 38
0136 and Polish Astroparticle Network 621/E-78/BWSN-0068/2008.
\end{acknowledgements}

\bibliographystyle{aa}

\bigskip
\bigskip
\noindent {\bf DISCUSSION}

\bigskip
\noindent {\bf ANDRZEJ ZDZIARSKI:} What is the current range 
of black hole masses in binaries and AGN?

\bigskip
\noindent {\bf BOZENA CZERNY:} The smallest one currently measured is 3.8 solar masses in object J1650-500 and comes from the paper by Shaposhnikov \& Titarchuk. As for the heaviest one, they are among the most distant quasars. For example, the black hole mass of $z = 6.41$ quasar SDSS J1148+5251 was estimated to be $3 \times 10^9 M_{\odot}$ \citep{willot}, and some quasars in Verstergaard (2009) sample have black hole masses up to 3 $\times 10^{10} M_{\odot}$. The black hole in OJ 287 mentioned in the context of binary black holes, with its mass of $\sim 1.8 \times 10^{10} M_{\odot}$ is also among the largest ones.

\bigskip
\noindent {\bf WOLFGANG KUNDT:} In Vestergaard et al. SDSS plot 
of the quasars as a function of redshift z, their average masses 
evolve from a few times $10^9 M_{\odot}$, at $z <= 5$, down to a few 
$10^6 M_{\odot}$ at $z <=0.2$. How can you downsize this inverted 
evolution to their predicted growth via accretion?

\bigskip
\noindent {\bf BOZENA CZERNY:} The plot shows only very active galaxies. 
The apparent rise of the quasar black hole mass with the redshift in that plot, and in the fits by Labita et al. (2009) to the quasar mass as $\propto (1 + z)^{1.64}$ only shows that more massive local over-densities evolve faster, which is known in cosmology as anti-hierarchical evolution. Those large mass over-densities becomes quasars at high z, and later on pass to a quiet non-active stage, not showing up in AGN surveys. Smaller over-densities evolve slowly, they become active much later and never grow up to give large black hole masses. For example, Sgr A* is likely to have occasional more active stages than seen at present but will never become a billion solar mass black hole.

\end{document}